\documentclass[10pt]{article}

\pdfoutput=1

\begin{document}
\title{American Physical Society 63rd Annual DFD Meeting: Detonation waves via molecular dynamics simulations of reactive hard disks}
\author{Nick Sirmas and Matei Radulescu \\
\\\vspace{5pt} Department of Mechanical Engineering, \\ University of Ottawa, Ottawa, Ontario K1N 6N5, Canada}
\maketitle
\section* \
The study of detonations at the microscopic level are of particular interest in having a better understanding of the hydrodynamics of reactive media. Observing detonations at the microscopic level demonstrates the role that molecular fluctuations play in the formation of detonations.

	In order to study at this scale, a molecular dynamic approach is used by treating the molecules as colliding hard disks. This approach allows for the study of the dynamics of gases, liquids and granular fluids.

	In the current video, we have a hard disk molecular dynamics (MD) simulation to study the hydrodynamics at the reactive regime. The system is initiated with 25000 hard disks, 12500 type A disks and 12500 type B disks. If these collide at an impact velocity above a defined threshold, they react to become type C disks. Kinetic energy is added to these disks to simulate the heat release of the reaction.

	In order to initiate the detonation a shock is propogated through the medium once it reaches a quasi-equilibrium state. The shock is formed by a piston propogating through the medium. The kinetic energy gain across the shock causes the disks to react. These series of events cause a detonation wave to propogate through the reactive system of particles. A close-up view of the reactions demonstrates the reactions at the microscopic level. 
\\\\
The video showing this MD simulation is:
\\
\textit{Detonation waves via molecular dynamics simulations of reactive hard disks}
\\\\
This fluid dynamics video has been submitted to the\textit{ Gallery of Fluid Motion} for the \textit{American Physical Society 63rd Annual DFD meeting}.

\end{document}